\documentclass[twocolumn,floatfix,pra,showpacs,preprintnumbers,superscriptaddress,amsmath,amssymb,10pt,aps]{revtex4-2}

\usepackage{amsmath}
\usepackage[export]{adjustbox}
\usepackage{graphicx,subfigure,amsmath,bbm}
\usepackage{mathtools}
\usepackage{mathptmx}
\usepackage{subfigure}
\usepackage{psfrag,graphicx}
\usepackage{dcolumn}
\usepackage{amsmath,amssymb}
\usepackage{bm}
\usepackage{color}
\usepackage{latexsym}
\usepackage{epstopdf}
\usepackage{color}
\usepackage[english]{babel}
\usepackage{latexsym}
\usepackage{psfrag,graphicx}
\usepackage{subfigure}
\usepackage{amsmath}
\usepackage{amssymb}
\usepackage{amsfonts}
\usepackage{bm}
\usepackage{natbib}
\usepackage{epstopdf}
\usepackage{xcolor}
\usepackage{leftindex}

\DeclareMathOperator{\Tr}{Tr}
\usepackage{appendix}
\newcommand{\RNum}[1]{\uppercase\expandafter{\romannumeral #1\relax}}

\definecolor{mygrey}{gray}{0.35}
\definecolor{myblue}{rgb}{0.2,0.2,0.8}
\definecolor{myzard}{cmyk}{0,0,0.05,0}
\definecolor{mywhite}{rgb}{1,1,1}
\definecolor{mywhite}{rgb}{1,1,1}
\definecolor{myred}{rgb}{1,0.,0.3}

\usepackage[colorlinks=true,citecolor=myblue,linkcolor=myred]{hyperref}

\def\ba{\begin{align}}
\def\enda{\end{align}}
\def\bi{\begin{itemize}}
\def\ei{\end{itemize}}

\def\be{\begin{equation}}
\def\ee{\end{equation}}
\def\bea{\begin{eqnarray}}
\def\eea{\end{eqnarray}}
\def\bse{\begin{subequations}}
\def\ese{\end{subequations}}

 \newcommand{\ket}[1]{|#1\rangle}

\begin{document}
\title{Probing the ergodicity breaking transition via violations of random matrix theoretic predictions for local observables}
\def\correspondingauthor{\footnote{Corresponding author: pivanov@phys.uni-sofia.bg}}
\author{Venelin P. Pavlov}
\affiliation{Center for Quantum Technologies, Department of Physics, St. Kliment Ohridski University of Sofia, James Bourchier 5 blvd, 1164 Sofia, Bulgaria}
\author{Peter A. Ivanov}
\affiliation{Center for Quantum Technologies, Department of Physics, St. Kliment Ohridski University of Sofia, James Bourchier 5 blvd, 1164 Sofia, Bulgaria}
\author{Diego Porras}
\affiliation{Institute of Fundamental Physics IFF-CSIC, Calle Serrano 113b, 28006 Madrid, Spain}
\author{Charlie Nation}
\affiliation{Department of Physics and Astronomy, University of Exeter, Stocker Road, Exeter EX4 4QL, United Kingdom}
\date{February 2025}

\begin{abstract}
 Quantum many-body systems can exhibit distinct regimes where dynamics is either ergodic, dynamically exploring an extensive region of available state-space, or non-ergodic, where the dynamics may be restricted. An example is the many-body localization (MBL) transition, where disorder induces non-ergodic behaviour. Most measures of ergodicity notably rely on global quantities, such as level spacing statistics. We explore the ability for a subsystem to probe the ergodicity of dynamics via measurement of local observables, and use expected results from random matrix theory (RMT) as a benchmark for the ergodic regime. We exploit two predictions from RMT as ergodicity is broken: the time evolution of the quantum Fisher information, and a fluctuation-dissipation relation. These are investigated in three different ergodicity breaking mechanisms, namely, as a consequence of transition to integrability, MBL, and Quantum Many-Body Scars (QMBS). We show that the predicted behaviour from RMT can be used as a potential witness for transition to non-ergodic behaviour from the measurement of local observables alone.
\end{abstract}

\maketitle

\section{Introduction} Typically systems in nature tend to thermalize in the sense that they fail to retain local information about their initial states. The expectation values of local observables in thermalizing systems at long times are equal to their microcanonical averages as assumed by the Eigenstate Thermalization Hypothesis (ETH) \cite{Deutsch1991,Deutsch2018,Srednicki1994,Srednicki1996}. However, in some cases ergodicity can be broken and the behaviour of local observables at long times deviate from the ETH prediction. For example this can be seen in Anderson localized systems, where localization arises in the presence of a disorder \cite{Anderson1958}. If an interaction is added between the particles and localization still persists, then the quantum system exhibits a phenomenon known as many-body localization (MBL) \cite{Fleishman1980,Abanin2019,Nandkishore2015,Gornyi2005,Basko2006,Znidaric2008,Pal2010}. Recently, breaking the ergodicity due to the MBL has been observed experimentally in various quantum-optical systems including for example cold atomic gases \cite{Lukin2019, Schreiber2015}, superconducting qubit array \cite{Gong2021}, and ion traps \cite{Smith2016}. 

Another mechanism to break ergodicity can be observed in systems which possess Quantum Many-Body Scars, ETH-violating eigenstates which may occur due to symmetries or near symmetries in an otherwise thermal spectrum \cite{Serbyn2021,Moudgalya2022,Chandran2023}. Experiments with Rydberg atoms have shown persistent long-time oscillations of local observables, which is ascribed to existing evenly spaced towers of QMBS \cite{Bernien2017}. These states are characterized by sub-volume-law entanglement entropy in contrast to the typical volume-law entanglement entropy of thermal states, which further demonstrates their ETH-violating property \cite{Iadecola2020}.

We note that the classical notion of ergodicity in the sense of any initial state exploring the entirety of phase space has no clear analogue in the quantum case, and various notions of quantum ergodicity can be described \cite{Shnirelman1974,Colin1985,Sunada1997,Zelditch1987,Zelditch1990}. In typical examples of the ergodic to non-ergodic transition, however, upon altering some parameter dynamics is dictated by a chaotic Hamiltonian in the ergodic regime, or an integrable Hamiltonian in the non-ergodic regime. For example in MBL, upon large enough disorder there is an effective description of the dynamics in terms of an integrable `$l$-bit' picture \cite{Serbyn2013_1}.

One of the most common measures for ergodicity thus arises from quantum chaos, the level-spacing statistics, which shows a Wigner-Dyson distribution in ergodic systems as predicted by RMT and Poissonian distribution in non-ergodic systems \cite{Alessio}. Another popular measure, especially in the MBL case, is the von-Neumann entanglement entropy $S$ which obeys a volume-law scaling for ergodic eigenstates \cite{Page1993}, a sub-volume law for QMBS \cite{Iadecola2020}, and an area-law for MBL eigenstates \cite{Bauer2013}. The entanglement growth also behaves differently. In ergodic systems it exhibits a ballistic growth \cite{Kim2013} and in MBL systems - a logarithmic growth, accompanied by a slow growth of the number entropy $S_N \sim \ln \ln t$ \cite{Bardarson2012, Kiefer2020}. 

As experimental capability to directly observe increasingly complex quantum dynamics progresses, observation of ergodicity breaking in many-body systems has seen substantial experimental interest \cite{Bordia2017, Saccone2023, AustinHarris2025}.
%
In many cases comparisons are made to RMT predictions for global quantities such as those mentioned above, however RMT methods for local observables \cite{Nation2018, Nation2019, Dabelow2020, Pavlov2024} may allow comparison to more easily measured quantities experimentally. In this work we see that analytical predictions made by an RMT approach for local observables yield accurate predictions for experimentally measurable quantities in the ergodic regime for multiple examples, which are violated in the non-ergodic regime, yielding potential experimental probes of ergodicity. Notably, the OTOC for local observables is known to probe ergodicity and the `full' ETH \cite{Alves2025probesoffull}, leaving an open question if simpler observable quantities can provide a similar ergodicity diagnostic.

In this work we numerically study multiple ergodicity breaking systems via coupling to a `probe' subsystem, for which we analyze local observable predictions from RMT. We first analyze the time-evolution of the quantum Fisher information (QFI), and the second is a relation between observable steady-state fluctuations and the decay rate to equilibrium. These are studied in three different scenarios: (i) A non-integrable-to-integrable spin-chain model transition, (ii) MBL in a disordered non-integrable spin chain model and (iii) QMBS in the PXP model \cite{Lesanovsky2012}, with an additional $B$-field acting on a single probe spin of the chain. 


The paper is organized as follows: In Sec. \ref{sec:setup} we introduce the general setting of our numerical results, and briefly describe the measures of ergodicity.  In Sec. \ref{Thermal} we briefly summarize the dynamical behaviour of the QFI in quantum ergodic systems and discuss the emergence of an intermediate linear regime from RMT predictions. In Sec. \ref{FlucsSection} we introduce the long-time fluctuations in ergodic systems and discuss their dependence on the shape of the eigenstates. In Sec. \ref{Breaking} we show our main investigation of three different ergodicity breaking mechanisms - Sec. \ref{sec:integrable} describes a non-integrable-to-integrable model transition, Sec. \ref{sec:MBL} a dynamical phase transition from a thermal phase to an MBL phase and Sec. \ref{sec:scars} QMBS. We show that in all three cases both the QFI and the scaling of long-time fluctuations of local observables efficiently captures the ergodicity breaking transition. Finally the conclusions are presented in Sec. \ref{Conclusion}.

\section{Set-up}\label{sec:setup}

In each of the studied systems below, we have a Hamiltonian of the form:
\begin{align}
    \hat{H} = \hat{H}_{\rm probe} + \hat{H}_{\rm bulk} + \hat{H}_{\rm coupling},
\end{align}
where we exploit RMT predictions on local observables of the probe Hamiltonian, chosen to be a single spin, in order to witness the ergodicity breaking transition in the total system. Crucially, in the ergodic regime, local observables of the probe system thermalize under effective RMT dynamics, where the probe is treated as a `system' with information lost to the `bath' Hamiltonian of the bulk. The probe observable in the ergodic regime this has statistically predictable scalings with tunable parameters such as coupling strength or system size. 
As mentioned above, we focus on two such predictions from RMT relating to the probe system, the QFI dynamics, and long-time observable fluctuations.

 
\subsection{Quantum Fisher Information in ergodic phase} \label{Thermal}

Let us consider a parameter-dependent quantum system described by a non-integrable Hamiltonian $\hat{H}_{\lambda}=\hat{H}_{0}+\hat{H}_{I}$, consisting of non-interacting Hamiltonian $\hat{H}_{0}=\hat{H}_{\rm S}+\hat{H}_{\rm B}$ with $\hat{H}_{\rm S}$ and $\hat{H}_{\rm B}$ being the Hamiltonians for the subsystem and the many-body environment respectively, and interaction part $\hat{H}_{I}$ describing the system-bath interaction. The system is initially prepared in an out-of-equilibrium state $|\psi_{0}\rangle$ which evolves under the action of unitary evolution, $|\psi_{\lambda}\rangle=\hat{U}_{\lambda}(t)|\psi_{0}\rangle$ with $\hat{U}_{\lambda}(t)=e^{-i \hat{H}_{\lambda}t}$. Then, the QFI is given by \cite{Paris2009}
\begin{equation}
F_{Q}(\lambda,t)=4(\langle\partial_{\lambda}\psi_{\lambda}|\partial_{\lambda}\psi_{\lambda}\rangle-|\langle\psi_{\lambda}|\partial_{\lambda}\psi_{\lambda}\rangle|^{2}). 
\end{equation}
The QFI has a simple geometrical interpretation as a measure of distinguishability between two infinitesimally close quantum states. It is also related to the experimentally accessible Loschmidt echo which measures the fidelity $F_{\epsilon}(t)=|\langle\psi_{0}|\hat{U}^{\dag}_{\lambda}(t)\hat{U}_{\lambda+\epsilon}(t)|\psi_{0}\rangle|^{2}$ between a state propagated forwards with $\hat{U}_{\lambda}(t)$ and propagated backwards with a slightly perturbed unitary operator $\hat{U}_{\lambda+\epsilon}(t)$, such that we have $F_{Q}(\lambda,t)=\lim_{\epsilon\rightarrow 0}\frac{1-F_{\epsilon}(t)}{\epsilon^{2}}$ \cite{Goussev2012,Gorin2006}.  

It is convenient to re-express the QFI in the eigenbasis of $\hat{H}_{\lambda}$, namely $\hat{H}_{\lambda}|\psi_{\mu}\rangle=E_{\mu}|\psi_{\mu}\rangle$. Assuming that the information for the parameter $\lambda$ is encoded in $\hat{H}_{0}$ we have
    \begin{eqnarray}
F_{Q}(\lambda,t)&=&4t^{2}\{\sum_{\mu\nu\rho}a^{*}_{\mu}a_{\nu}O_{\mu\rho}O_{\rho\nu}
e^{i\theta_{\mu\nu}t}{\rm sinc}(\theta_{\mu\rho}t)
{\rm sinc}(\theta_{\rho\nu}t)\notag\\
&&-|\sum_{\mu\nu}a^{*}_{\mu}a_{\nu}e^{i\theta_{\mu\nu}t}O_{\mu\nu}{\rm sinc}(\theta_{\mu\nu}t)|^{2}\},\label{QFI_1}
\end{eqnarray}
where $O_{\mu\nu}=\langle\psi_{\mu}|\partial_{\lambda}\hat{H}_{0}|\psi_{\nu}\rangle$ are the matrix elements in the many-body interacting basis of the observable $\hat{O}$, $\theta_{\mu\nu}=(E_{\mu}-E_{\nu})/2$, $a_{\mu}=\langle\psi_{\mu}|\psi_{0}\rangle$, and ${\rm sinc}(x)=\sin(x)/x$. The expression (\ref{QFI_1}) is suitable for defining the short and long time behaviour of QFI. Indeed, using that $\lim_{x\rightarrow 0}{\rm sinc}(x)=1$ one can obtain the short time behaviour of the QFI, namely $\lim_{t\rightarrow 0}F_{\rm Q}(\lambda,t)/t^{2}={\rm const}$, indicating that it grows quadratically with time. Similarly, we may consider the long-time limit of the QFI by using that $\lim_{x\rightarrow \infty}{\rm sinc}(x)=0$, where again the QFI time scaling is quadratic, $\lim_{t\rightarrow \infty}F_{\rm Q}(\lambda,t)/t^{2}={\rm const}$. Recently, we showed that for quantum ergodic systems there exists an intermediate linear with time scaling of the QFI, $F_{\rm Q}(\lambda,t)\sim t$, which follows directly from RMT calculation of (\ref{QFI_1}) \cite{Pavlov2024}. The crossover between the linear-to-long time quadratic time regimes occurs at the Heisenberg time $\tau\approx D(E)$ where $D(E)$ is the density of states at the initial state energy $E=\langle\psi_0|\hat{H}|\psi_0\rangle$. We expect that any processes that break the ergodicity of the system would reduce the effective density of states, describing those states dynamically explored from a typical initial state. Consequently, this will affect the parameter information flow in terms of the QFI dynamics, which will exhibit a standard quadratic growth \cite{Paris2009}. Therefore the vanishing of the linear time growth of the QFI, may serve as a figure of merit for ergodicity breaking in a quantum system. 
\begin{figure*}[!htbp]
\hspace*{-0.0cm}
\centering
    \includegraphics[width=1\textwidth,trim=0.3cm 0 0 0,
  clip]{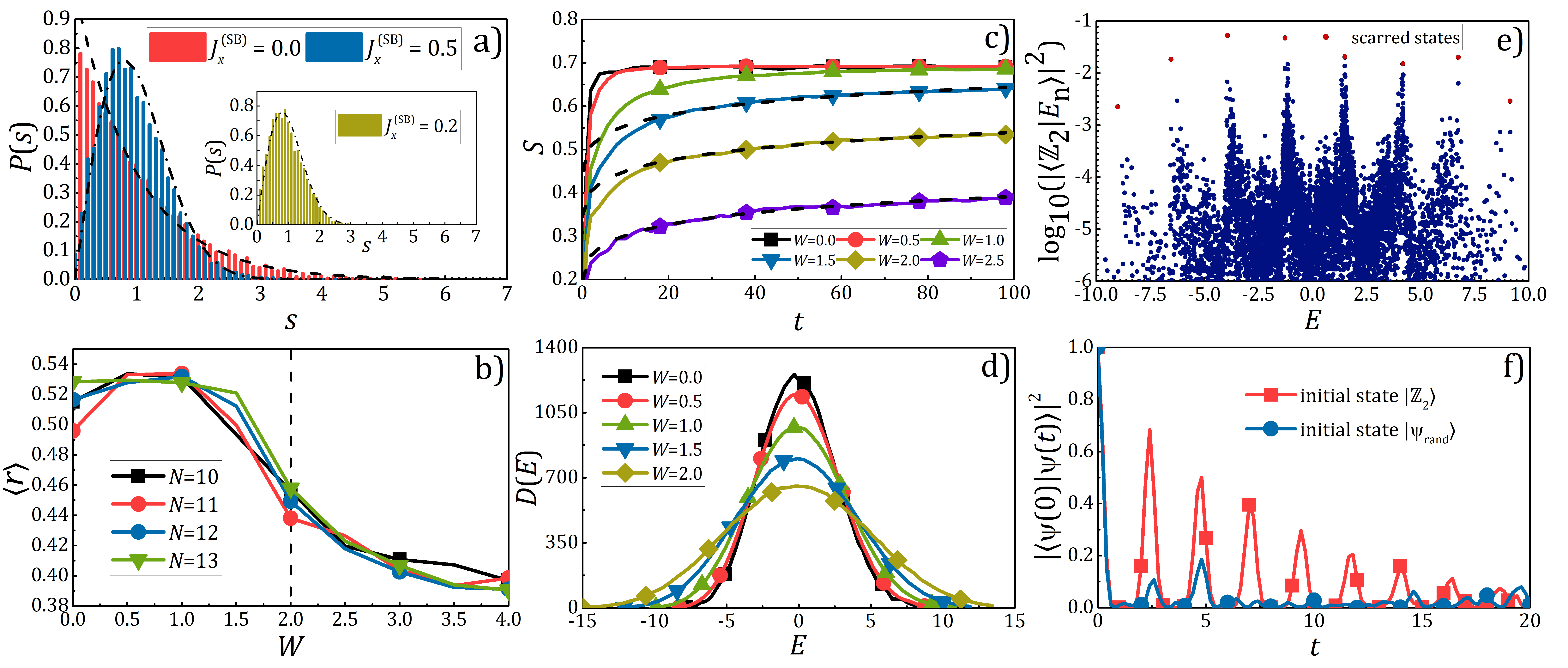}
    \caption{Standard measures for quantum ergodicity. a) Energy level-spacing statistics $P(s)$ for $N=13$ and different coupling constants $J_x^{(SB)}$. The dashed line denotes Poisson distribution $P(s)=e^{-s}$ and the dash-dotted line denotes Wigner-Dyson distribution $P(s)=\frac{\pi}{2} se^{-\pi s^2/4}$. The parameters are set to $B=0.01$, $B_x^{(\rm B)}=0.3$, $J_z^{(\rm SB)}=0.2$, $J_x =1$. b) The average ratio of consecutive energy spacings $\langle r\rangle$ as a function of the disorder $W$ for different number of spins $N$. The coupling constant is set to $J_x^{(\rm SB)}=0.4$. The results for $W \neq 0$ are averaged over 30 iterations. c) Time evolution of the von-Neumann entropy $S$ for $N=11$ and initial state $|\Psi_0\rangle = |\varphi_{\alpha_0} \rangle$ with $\alpha_0=1000$ and various disorder strengths $W$. The numerical results are compared with the function $a\ln(t)+b$ (dashed lines). The results for $W \neq 0$ are averaged over 30 iterations. d) The density of states $D(E)$ as a function of the disorder $W$ for $N=13$. The results for $W \neq 0$ are averaged over 30 iterations. e) Overlap of $|\mathbb{Z}_2\rangle$ with the eigenstates of $\hat{H}_{\rm QMBS}$ as a function of the energies $E$ for $B=0.4$ and number of spins $N=20$. The scarred states are denoted in red. f) The survival probability as a function of time for different initial states. Revivals can be seen for $|\mathbb{Z}_2\rangle$.}\label{fig1}
\end{figure*}

\subsection{Long-time fluctuations in the ergodic phase}\label{FlucsSection}

In the ergodic phase, for an initial non-equilibrium state, local observables decay to a thermal expectation value over some timescale $\Gamma^{-1}$, and fluctuate in time. Specifically, the long-time fluctuations of an observable $\hat{O}$, may be quantified via
\begin{eqnarray}
    \delta_{\hat{O}}^2(\infty) = \lim_{T\to\infty} \frac{1}{T} \int_0^T dt (\langle \hat{O}^2(t)\rangle - \langle \hat{O}(t)\rangle^2).
\end{eqnarray}
These fluctuations can be directly calculated via RMT, and related to the observable decay rate via a fluctuation-dissipation relation, derived in Ref. \cite{Nation2019}:
\begin{equation}\label{eq:QCFDT}
    \delta^2_{\hat{O}}(\infty) = \chi \frac{\overline{[\Delta O^2]}}{4\pi D(E) \Gamma},
\end{equation}
where $\overline{[\Delta O^2]} := \overline{[O^2]} -\overline{[O]}^2$, with $\overline{[O]} := \sum_\mu |c_\mu(\alpha)|^2 O_{\alpha\alpha}$. We can then use the prediction that $\delta^2_{\hat{O}}(\infty) \sim \frac{1}{D(E) \Gamma}$ as a fingerprint of the ergodic regime. Notably, the $1 / 4\pi$ prefactor originates from the assumption that the shape of coarse-grained eigenstates takes a Lorentzian form, which is typical for chaotic models in weak coupling regimes \cite{Nation2018}, however the numerical value of the prefactor can alter for differing shapes \cite{ChorbadzhiyskaIvanovNation2025MultiTimeCorrelations}. We thus introduce the parameter $\chi$ to which may take a different value depending on the model. This will be seen in the PXP example below, which is the only case studied where $\chi \neq 1$ due to deviation from Lorentzian eigenstates. In order for the measure to be further model-agnostic one can note that as $D(E) \sim 2^N$ up to some constant in the chaotic regime for $E$ close to the middle of the spectrum \cite{Nation2019_1}, the scaling can be obtained from only measurable quantities $\delta^2_{\hat{O}}(\infty)$ and $\Gamma$.

\section{Results} \label{Breaking}


In this section we will introduce three models which each have an ergodicity breaking transition with a different mechanism. We will show that the two measures introduced above, the QFI and long-time fluctuations, each act as witnesses of the transition - obeying the RMT results in the ergodic regime, with a clear violation as ergodicity is broken.

\begin{figure*}[!htbp]
\hspace{-0.0cm}
\centering              
    \includegraphics[width=1.0\linewidth,trim=0.0cm 0 0 0, clip]{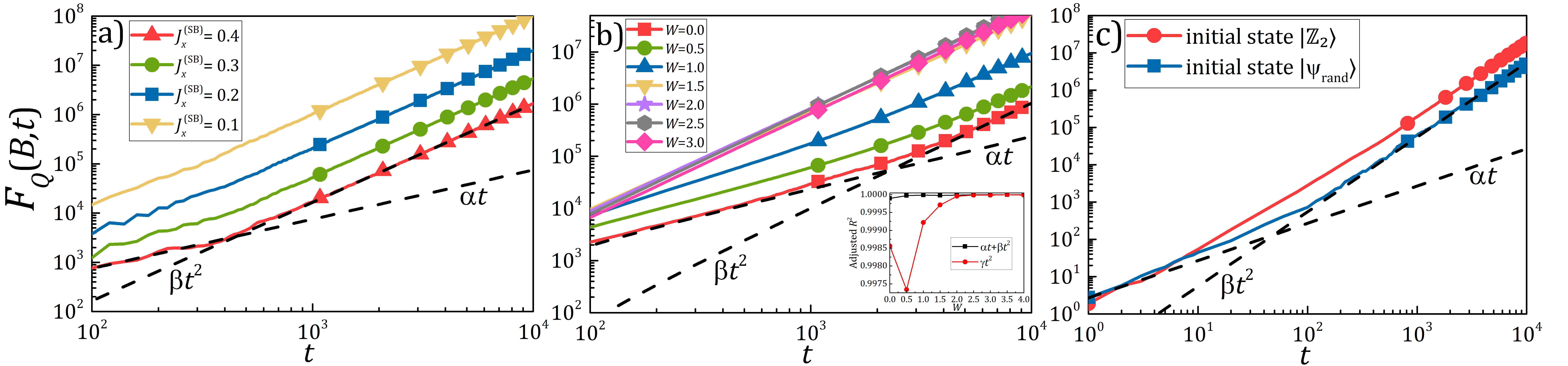}
    \caption{Comparison of the QFI in all three ergodicity breaking mechanisms. a) Long time evolution of the QFI for different coupling strengths $J^{(\rm SB)}_{x}$ for $N=11$ and initial state $|\Psi_0\rangle = |\uparrow_x\rangle \bigotimes|\psi_{\rm B} \rangle$ where $|\psi_{\rm B} \rangle$ is an eigenstate of $\hat{H}_{\rm B}$. The dashed lines show the functions $\alpha t$ and $\beta t^2$ for $\alpha=7.5$ and $\beta=0.0165$. The parameters are set to $B=0.01$, $B_x^{(\rm B)}=0.3$, $J_z^{(\rm SB)}=0.2$, and $J_x =1$ b) Long time evolution of the QFI for various disorders $W$ for $N=13$ and initial state $|\Psi_0\rangle = |\varphi_{\alpha_0} \rangle$ with ${\alpha_0}=5500$. The dashed lines show the functions $\alpha t$ and $\beta t^2$ for $\alpha=26$ and $\beta=0.024$. The results for $W \neq 0$ is averaged over 30 iterations. (inset) The adjusted coefficient of determination $R^2$ for the fits $\alpha t + \beta t^2$ and $\gamma t^2$ as a function of the disorder $W$. The vertical dashed line shows the dynamical phase transition point $W_c$. c) Long time evolution of the QFI for different initial states, for $N=16$ and $B=0.4$. The dashed lines show the functions $\alpha t$ and $\beta t^2$ for $\alpha=2.7$ and $\beta=0.1$.}
\label{fig2}
\end{figure*}
The first two ergodicity breaking mechanisms each begin from considering 1D spin system with a Hamiltonian of the form

\begin{equation}
\hat{H}=\hat{H}_{\rm S}+\hat{H}_{\rm B}+\hat{H}_{\rm SB}.\label{spinH}
\end{equation}
The system Hamiltonian describes a single probe spin in the presence of a $B$-field
\begin{equation}
\hat{H}_{\rm S}=B\sigma^{z}_{1},
\end{equation}
where $\sigma^{q}_{j}$ ($q=x,y,z$) are the Pauli matrices acting on $j$-th site. The bath Hamiltonian describes a spin chain with Ising interaction
\begin{equation}
\hat{H}_{\rm B}=\sum_{k>1}^{N} B_x^{(\rm B)}\sigma_k^{x} + \sum_{k>1}^{N-1}J_x(\sigma^{+}_{k}\sigma^{-}_{k+1}+\sigma^{-}_{k}\sigma^{+}_{k+1}),
\end{equation}
where $B^{(B)}_{x}$ is the magnetic field along the $x$-axis and $J_x>0$
is the spin-spin coupling. The interaction Hamiltonian describes a coupling between the system spin and a single bath spin of index $r$ 
\begin{equation}\label{eq:coupling}
\hat{H}_{\rm SB}=J_z^{(\rm SB)}\sigma^{z}_{1}\sigma^{z}_{r}+J_x^{(\rm SB)}(\sigma^{+}_{1}\sigma^{-}_{r}+\sigma^{-}_{1}\sigma^{+}_{r}),
\end{equation}
with coupling strengths $J^{(\rm SB)}_{z}$ and $J^{(\rm SB)}_{x}$.

\subsection{Non-Integrable-To-Integrable Model Transition}\label{sec:integrable}
In this subsection we will illustrate the simplest way to break ergodicity in the spin system (\ref{spinH}). To do so, we note that the chain $\hat{H}_B$ is itself integrable, and non-integrability can be induced by the probe subsystem itself by coupling to a mid-chain spin as in e.g \cite{Nation2019}. Here we choose $r = 5$ in Eq.~\eqref{eq:coupling}. We thus compare extremely weak couplings $J^{(\rm SB)}_{x}$ between the probe system and bath, which are dominated by integrable behaviour, to stronger couplings where chaoticity of the system ensures ergodic dynamics. This can be seen in Fig. \ref{fig1}a) where we show the energy level-spacing statistics for varying $J^{(\rm SB)}_{x}$, and observe a crossover from Poissonian to Wigner-Dyson statistics, indicating the onset of chaoticity and effective RMT behaviour, at $J^{(\rm SB)}_{x} \approx 0.2$. 

\begin{figure*}[!htbp]
\hspace*{-0.1cm}
\centering
    \includegraphics[width=1.02\linewidth,trim=0.3cm 0 0 0,
  clip]{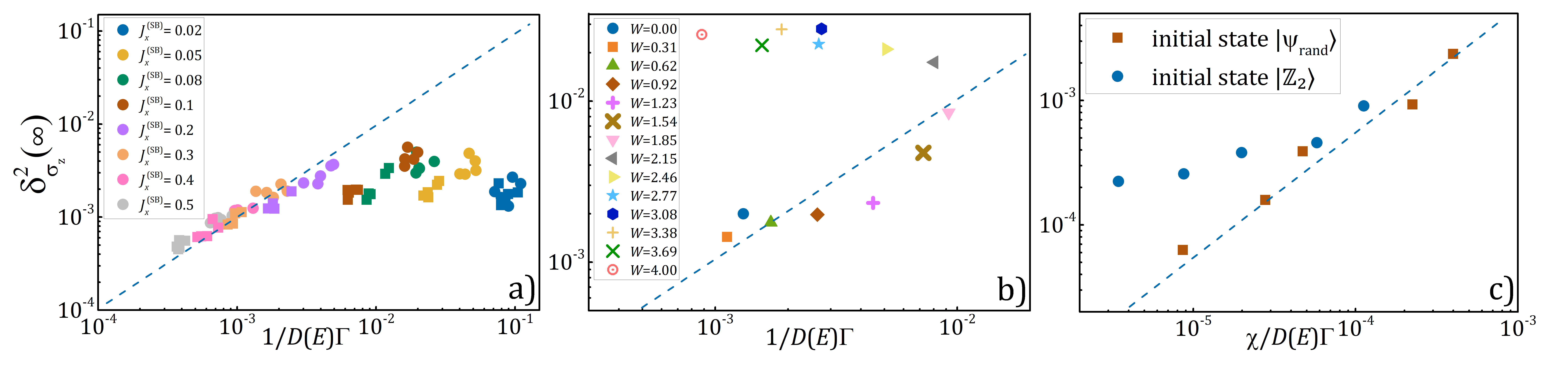}
    \caption{a) Shows violation of fluctuation-dissipation relation (labeled RMT) for weak couplings in the integrable-ergodic transition. Circles show $N=13$, squares show $N=14$. We see that for both system sizes deviation from the RMT prediction occurs for couplings $J_z \lesssim 0.2$. The dashed line denotes the RMT result. b) Shows the fluctuation-dissipation relation for the MBL transition rescaled by constants as in Eq. \eqref{eq:QCFDT}. Here we similarly see a deviation from the RMT prediction for $W \gtrsim 2$, as observed for ergodicity measures in Fig. \ref{fig2}. c) Shows fluctuation-dissipation relation for PXP model for $B = 0$ and a central probe spin for varying $N =[14, 16, 18, 20, 22] $ (increasing from right to left) for both scarred and ergodic initial states.}
\label{fig3}
\end{figure*}

In order to demonstrate the use of RMT predictions for witnessing this breaking of ergodicity for weak coupling strengths, in Fig. \ref{fig2}a) we show the time evolution of the QFI for $\lambda=B$ and for different values of the coupling parameter $J^{(\rm SB)}_{x}$. Notably, decreasing the strength of the interaction between the subsystem and the bath causes a reduction of the linear time growth of the QFI which eventually vanishes as the coupling $J^{(\rm SB)}_{x}$ becomes smaller and the model approaches the integrable limit. Similarly, in Fig. \ref{fig3}a), we show the long-time fluctuations of the observable $\sigma_1^z$ for varying $J^{(\rm SB)}_{x}$, which tell a similar story: for large $J^{(\rm SB)}_{x}$ the observed fluctuations scale with Eq. \eqref{eq:QCFDT}, as expected from RMT, however as the coupling is decreased, and the dynamics becomes quasi-integrable due to the integrability of the bath Hamiltonian, the scaling is violated, and ergodicity breaking is witnessed. Crucially, all three measures, level statistics, QFI, and long-time fluctuations, observe this transition at $J^{(\rm SB)}_{x} \approx 0.2$. 

\subsection{Many-Body Localized phase}\label{sec:MBL}
A further example of an ergodicity breaking transition is the emergence of many-body localization in strongly disordered systems. Here we consider the spin system described by Eq. (\ref{spinH}) and add a disorder term $\hat{H}_{\rm D}$, such that
\begin{equation}
\hat{H}=\hat{H}_{\rm S}+\hat{H}_{\rm B}+\hat{H}_{\rm SB}+\hat{H}_{\rm D}\label{H_D}.    
\end{equation}
The disorder Hamiltonian describes a site-dependent disordered potential
\begin{equation}
\hat{H}_{\rm D} = \sum_{i=1}^{N} D_i \sigma^z_i,
\end{equation}
where $D_i$ is taken from a uniform random distribution $D_i \in [-W,W]$. We choose the system-bath coupling strengths such that for $W = 0$ the system is in the ergodic regime, which can be confirmed by the average ratio of consecutive level spacings Fig. \ref{fig1}b) as we discuss below. After a disordered potential is applied to the system, for suitably large $W$ a localisation transition takes place, and the dynamics is no longer ergodic.

A common marker of this transition to the MBL phase is the dynamics of the von-Neumann entropy $S = - \Tr(\hat{\rho}_{\rm S}\ln\hat{\rho}_{\rm S})$, which quantifies the entanglement between the subsystem and the bath, with $\hat{\rho}_{\rm S}={\rm Tr}_{\rm B}\hat{\rho}$ being the reduced density matrix. Moreover in the localized regime one can write the Hamiltonian in the form \cite{Huse2014}:
\begin{equation}
\hat{H}_{{\rm MBL}}=\sum_i h_i {\tau_i^z} + \sum_{i,j} J_{ij}{\tau_i^z}{\tau_j^z}+\sum_{i,j,k} J_{ijk}{\tau_i^z}{\tau_j^z}{\tau_k^z} \; + \;...,
\end{equation}
where the set of operators ${\{\tau_i\}}$ are called pseudospins or l-bits and can be obtained from the Pauli operators ${\sigma_{i}^{\alpha}}$ by means of a quasilocal rotation $\Omega$, such that $\tau_i^z = \Omega\sigma_{i}^{\alpha}\Omega^{\dagger}$ \cite{Imbrie2016}. These operators commute with each other and with the Hamiltonian and therefore form a complete set of local integrals of motion. The magnitude of the interactions decays exponentially with the separation between the pseudospins $J_{i_1,...i_k} \propto e^{-\max|i_\alpha - i_\beta|/\xi}$, where $\xi$ is a characteristic length scale. In Fig. \ref{fig1}c) it can be seen that the disorder leads to a logarithmic growth as the system becomes more localized. For two initially not entangled l-bits, this interaction entangles them after time $t$ such that $J_{ij}t\geq 1$. From here it follows that the entanglement between them generates within a distance $|i-j| \sim \xi\ln(J_{ij}t)$, which explains the logarithmic entropy growth \cite{Imbrie2017,Ros2015,Serbyn2013_2}.

In order to find the approximate transition point $W_c$ between the thermal and MBL phases we turn our attention to the spectral statistics of adjacent energy levels of the Hamiltonian (\ref{H_D}). In the ETH phase the level spacing $s_n = E_{n}-E_{n-1}$ follows a Wigner-Dyson distribution as predicted by RMT. In localizing systems however, the correlations between the eigenvalues become disrupted, which leads to Poisson distribution of the spacing. A convenient figure of merit to distinguish between both regimes is the ratio of consecutive energy spacings $r_n = \frac{{\rm min}(s_n,s_{n-1})}{{\rm max}(s_n,s_{n-1})}$ \cite{Oganesyan2007}. For Wigner-Dyson statistics it's average value is given by $\langle r \rangle_{{\rm WD}} \approx 0.53$, whereas for Poisson statistics it is given by $\langle r \rangle_{{\rm Poisson}} \approx 0.39$ \cite{Atas2013}. In Fig. \ref{fig1}b) we show the average value $\langle r\rangle$ for various $N$. As the number of spins is increased the ratio becomes steeper at around $W_c\sim2$, and in the thermodynamic limit $N\rightarrow \infty$ approaches a step function. The critical value $W_{c}$ marks the phase transition point between thermal and MBL phases. It is important to emphasize that the entropy dynamics in Fig. \ref{fig1}c) is not able to fairly capture the transition point to the MBL phase visually in contrast to the ratio of consecutive energy spacings.


Next we investigate the behaviour of the QFI in the MBL phase. In Fig. \ref{fig2}b) we see that the QFI grows as the disorder is increased and reaches a maximal value close to the critical point $W_c$. One can also see that a distinctive property of the QFI that indicates the dynamical phase transition is the vanishing of the linear regime, which can be seen to completely vanish at around $W_c \sim 2$. In the thermal phase for small disorder strengths the Heisenberg time $\tau$ emerges earlier in the evolution, shortening the timespan of the linear regime. This can be explained by the change of the density of states at the initial energy $D(E)$ to which the Heisenberg time is proportional \cite{Pavlov2024,Schiulaz2019}. While the system is still in a thermal phase for not so large disorder we can assume that the general RMT expression is still valid. If we take a look at Fig. \ref{fig1}d) we see that $D(E)$ decreases as the disorder increases, which explains the behaviour of the crossover time between both regimes. As seen in Fig. \ref{fig2}b) the linear regime is completely gone in the MBL phase and the QFI decreases for a given time when increasing $W$ beyond the transition point, indicating a reduction in entanglement in the system.

In order to statistically display the absence of a linear regime in the MBL phase, we investigate the adjusted coefficient of determination $R^2$ as a measure of how well observable data is explained by a model and whether adding an additional term to the fit is helpful \cite{Ezekiel1930,Burnham2002,Glantz2016}. Here we use it to show that inclusion of the linear term in a fit to $\alpha t+\beta t^2$ doesn't significantly improve the model's explanatory ability in the MBL phase. In Fig. \ref{fig2}b) (inset) we plot the adjusted $R^2$ as a function of the disorder $W$ for both fits. We see that both functions become equal at the critical point $W_c \sim 2$, suggesting that the linear regime becomes statistically redundant in the MBL phase, and a quadratic function $\gamma t^2$ is sufficient to describe the dynamical behavior of the QFI.

In Fig. \ref{fig3}b) we show the scaling of long time fluctuations of the observable $\sigma_1^z$ with $W$, where we similarly observe an abrupt change in behaviour for $W > 2$. Prior to this value, the fluctuations approximately follow the RMT prediction. Notably, small deviations are expected due to the sensitivity of the numerical prefactor to the specific shape of chaotic eigenstates, which can itself be affected by $W$. After the value $W \sim 2 $, a distinct transition is observed, where the fluctuations $\delta^2_{\sigma_z}(\infty)$ remain approximately constant, rather than decreasing with $W$, indicating the localisation transition. This behaviour can be intuitively understood, as long-time fluctuations in the ergodic regime decrease with $D(E)$, which can be understood as an effective number of eigenstates of $\hat{H}_0$ and eigenstate at energy $E$ overlaps with (indeed the inverse participation ratio (IPR) can be seen to scale similarly \cite{Nation2018}). 

\begin{figure}[!htbp]
\hspace*{-0.5cm}
\includegraphics[width=0.47\textwidth]{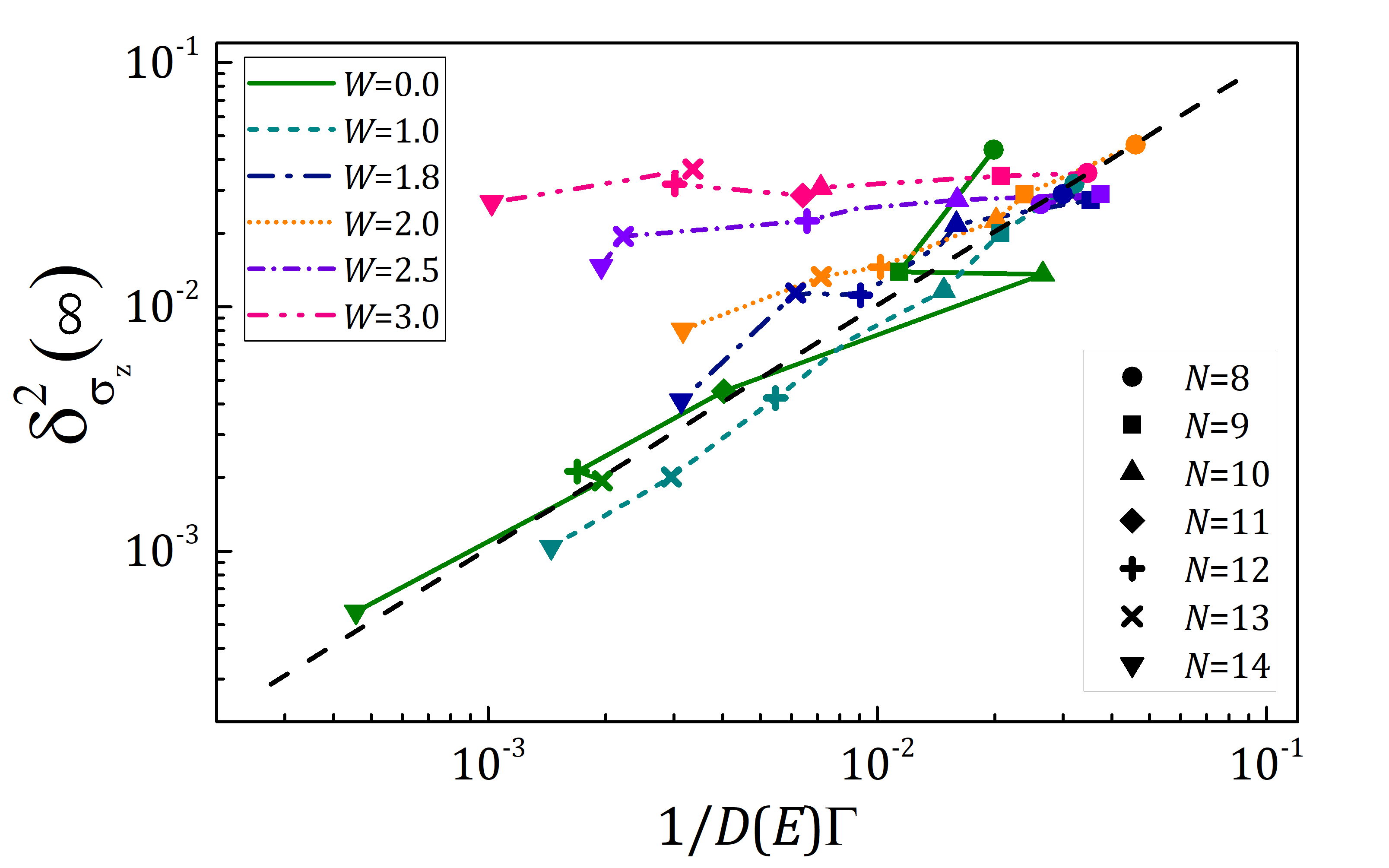}
\caption{Shows fluctuation-dissipation relation for the MBL transition for different disorder strengths $W$ and various numbers of spins $N$.}
\label{fig4}
\end{figure} 
In the ergodic regime, this exponential decrease in fluctuations occurs due to the initial state exploring the entirety of the available Hilbert space. After the localisation transition, however, there is a finite length scale along which an initial perturbation may propagate, and the effective Hilbert space dimension is no longer extensive in system size.

This can be further seen in Fig. \ref{fig4}, where we show the scaling of fluctuations with system size for different disorder strengths $W$. In this case, below the localisation transition we see that the fluctuations indeed scale with the fluctuation relation from RMT. At $W \sim 2$, however, we see a change in behaviour, with the fluctuations decreasing far slower than expected. For even larger $W$ the fluctuations stay constant when $N$ is decreased beyond some value, which we interpret as the chain increasing beyond the localisation length scale, after which increasing the length of the chain no longer provides a larger effective Hilbert space for the state to explore, and hence fluctuations no longer decrease.





\subsection{Quantum Many-Body Scars}\label{sec:scars}

Here we turn our attention to a different ergodicity breaking mechanism, that of QMBS. We first introduce the PXP model, which is an approximation of an experiment with Rydberg atoms under a Rydberg blockade constraint in which quantum scars were observed \cite{Bernien2017}. The Hamiltonian is given by:
\begin{equation}
    {\hat H}_{\rm PXP} = \sum_{i=1}^{N} P_{i-1}\sigma_i^xP_{i+1},
\end{equation}
where $P_i=(1-\sigma^z_i)/2$ are projector operators into the spin down state. The structure of the Hamiltonian ensures that two neighboring spins can not be in their excited states simultaneously.
In order to investigate the QFI, we add a local $B$-field to a single spin of the chain, so that we have:
\begin{equation}
    {\hat H}_{\rm QMBS} = B\sigma_1^z + {\hat H}_{\rm PXP}.
\end{equation}
In this way the first spin acts as our probe system in the PXP model. In our analysis we consider the open boundary conditions (OBC) case, in which the boundary terms are given by:
\begin{equation}
{\hat H}_{\rm OBC} = \sigma_1^x P_2 + P_{N-1}\sigma_N^x.
\end{equation}
This Hamiltonian shows prominent features, labeled QMBS, in the sense that although it is non-integrable and is expected to be thermalizing for generic states, it hosts a sparse set of atypical eigenstates embedded in an otherwise thermal spectrum. These ``scarred'' eigenstates violate the ETH expectations by exhibiting anomalously low entanglement, and they have disproportionately large overlap with a small number of simple product states in the constrained Hilbert space. 

A useful way to view the constraint is to work in the $\sigma^z$ basis, where $\ket{\uparrow}$ denotes an excited (Rydberg) atom and $\ket{\downarrow}$ the ground state. The blockade condition forbids adjacent excitations, so the accessible Hilbert space consists of all binary strings with no neighboring $\uparrow\uparrow$ pairs, whose dimension grows only as the Fibonacci sequence rather than $2^N$. Within this constrained space, the dynamics starting from special charge-density-wave (N\'eel) states \cite{Bernien2017,Sun2008,Olmos2009},
\begin{equation}
\ket{\mathbb{Z}_2}=\ket{\uparrow\downarrow\uparrow\downarrow\cdots},\qquad 
\ket{\mathbb{Z}'_2}=\ket{\downarrow\uparrow\downarrow\uparrow\cdots},
\end{equation}
is particularly anomalous: local observables display long-lived, approximately periodic revivals instead of rapidly relaxing to stationary thermal values. Microscopically, these revivals are associated with an approximately equally spaced ``tower'' of scarred eigenstates that can be organized by an emergent, approximate $\mathrm{SU}(2)$ structure, which supports coherent oscillations between low-entanglement configurations. The overlap of eigenstates of $\hat{H}_{\rm QMBS}$ with the state $\ket{\mathbb{Z}_2}$ are shown in Fig. \ref{fig1}e), showing the structure of the scarred spectrum. We note that the addition of a probe term to the PXP does not significantly alter the structure of the scars (denoted in red), which are prominent features despite inclusion of the field $B$.

The effect of QMBS on the dynamics of a system can be clearly seen by analysis of the survival probability, given by:
\begin{equation}
F(t)=|\langle\psi(0)|\psi(t)\rangle|^2,
\end{equation}
which measures how much an evolved state overlaps with the initial state.  In Fig. \ref{fig1}f) we show the survival probability for two different initial states. The first, $|\mathbb{Z}_2\rangle$, has a high overlap with the scarred eigenstates, the second initial state is chosen as a random superposition of eigenstates $|\psi_{\rm rand}\rangle = \sum_{n=1}^{M} c_n|E_n\rangle$, which have small overlap with $|\mathbb{Z}_2\rangle$ i.e. the most thermal states.
It can be seen that there are significantly increased revivals from the $|\mathbb{Z}_2\rangle$ initial state, whereas choosing an initial state $|\psi_{\rm rand}\rangle$ from a random superposition of bulk eigenstates yields ergodic thermalizing dynamics.

In Fig. \ref{fig2}c) we plot the QFI dynamics for the same two initial states, where we can see that we have a purely quadratic scaling when using an initial state with high QMBS overlap, and the usual linear-to-quadratic scaling for an initial state built with the ETH-obeying eigenstates in the PXP model. This supports our previous observation, that the linear regime of the QFI is indicative of ergodic dynamics. We note that as the scaling prefactor of the fluctuation-dissipation relation depends crucially on the shape of the chaotic wavefunctions \cite{Nation2019}, the chaotic scaling is not necessarily observed if the shape of these wavefunctions alters between realizations of different $N$. We find that choosing the probe qubit to be central in the PXP chain minimises this effect.

Finally, in Fig. \ref{fig3}c) we show the fluctuation-dissipation relation for varying system size $N$ of the PXP model for the scarred and ergodic initial states. In this case we see distinctly differing scaling for the scarred state, again showing observably differing behaviour from that expected from RMT. Notably in this case the eigenstates of the PXP model are not of Lorentzian form, and thus the numerical prefactor $\chi$ in Eq. \eqref{eq:QCFDT} differs from one. The numerical value is found to be $\chi \sim 5.5$. Crucially, this fit requires no additional data, and thus determination of the ergodic scaling remains experimentally assessable even in such cases where the numerical prefactor is not analytically obtainable. We thus observe, as in the previous ergodicity breaking transitions, that ergodic dynamics in the PXP model may be witnessed by observable fluctuations.


\section{Conclusion} \label{Conclusion}

In this work we have shown that predictions from RMT for the behaviour of local observables of a probe system may be exploited to observe ergodicity breaking in quantum many-body systems. The central RMT predictions exploited here are a linear regime of growth of the QFI \cite{Pavlov2024}, and a fluctuation-dissipation relation \cite{Nation2019}. Each of which describe the behaviour of local observables in the ergodic regime. We have studied three models which each display ergodic and non-ergodic regimes, however in very different manifestations, and shown that in each case the fingerprint of the ergodic regime may be characterised by the RMT predictions, which are violated when ergodicity is broken.

The first, and conceptually simplest model studied, is a spin chain model where an integrable spin chain is perturbed by an out-of-plain spin, which breaks the integrability of the system. From observables of the perturbing spin, we see that for very weak couplings the RMT predictions do not match, indicating that the integrable nature of the spin chain dominates dynamics for weak couplings. As the coupling strength increases to a modest value, RMT predictions well capture the behaviour of both the QFI and long-time fluctuations.

The second model we treat is a non-integrable spin-chain model with local on-site disorder, which undergoes a MBL transition for strong enough disorder strenghts. The critical value can be estimated from the energy level-spacing statistics, which are Wigner-Dyson in the ergodic regime, and Poissonian in the localised (non-ergodic) regime. We find that both the QFI dynamics and long-time fluctuations observe this MBL transition at comparable disorder strengths to the level statistics.

The final model we analyse is the PXP model, which does not show distinct parameter regimes, but rather its eigenstates display either ergodic or non-ergodic behaviour, with the latter due to quantum scars in the spectrum. We show that for randomly selected initial states with small overlap with the quantum scarred states, the RMT predictions have good agreement with numerical calculations. For initial states with large overlap with the scarred states, the RMT predictions are violated.

Our work thus shows the broad validity of RMT predictions in ergodic quantum systems, as well as the potential for the violation of these predictions to be exploited to observe ergodicity breaking phase transitions via measurement of local observables.

\section*{Acknowledgments}
V. P. P. and P. A. I. acknowledge the Bulgarian national plan for recovery and resilience, contract BG-RRP-2.004-0008-C01 (SUMMIT: Sofia University Marking Momentum for Innovation and Technological Transfer), project number 3.1.4. 
C. N. acknowledges funding from the EPSRC quantum
career development grant EP/W028301/1 and the
EPSRC Standard Research grant EP/Z534250/1. D. P. acknowledges support from Spanish project PID2024-159152NB-I00.

\end{document}